\documentclass[12pt]{article}

\usepackage{epsfig}

\usepackage{amssymb}

\def\be{\begin{equation}}
\def\ee{\end{equation}}
\def\ba{\begin{array}{c}}
\def\ea{\end{array}}

\newcommand{\bea}{\begin{eqnarray}}
\newcommand{\eea}{\end{eqnarray}}

\newcommand{\pbr}{\prec\!}
\newcommand{\pkt}{\!\!\succ\,\,}
\newcommand{\kt}{\rangle}
\newcommand{\br}{\langle}

\newcommand{\bbbr}{\{}
\newcommand{\kkkt}{\}}

\begin{document}

\begin{center}

{\Large \bf

Hermitian--to--quasi-Hermitian quantum phase transitions

  }

\vspace{9mm}

{Miloslav Znojil}

\vspace{9mm}

Nuclear Physics Institute CAS, 250 68 \v{R}e\v{z}, Czech Republic

{znojil@ujf.cas.cz}

{http://gemma.ujf.cas.cz/$\sim$znojil/}

\end{center}

\vspace{9mm}


\section*{Abstract}

The phenomenon of quantum phase transition is considered in the
special case in which the evolution laws remain unitary and in which
the bound-state energies remain observable. The conventional
Hermiticity of observables is lost at the interface, replaced by the
so called quasi-Hermiticity. Several features of the passage of the
system through the interface are discussed and illustrated by
elementary illustrative ${\cal PT}-$symmetric examples.

\section*{Keywords}

unitary quantum theory; non-Hermitian representation;  evolution
equations; phase transitions; PT symmetric benchmark model;

\newpage

\section{Introduction\label{introdukce}}

In historical retrospective the birth of  quantum mechanics was
certainly facilitated by several remarkably friendly experimental as
well as theoretical aspects of its applicability, say, to hydrogen
atom or to elementary molecules \cite{Messiah}. Even the transition
from non-relativistic hydrogen-atom-like Hamiltonians $\mathfrak{h}$
to their relativistic Dirac-equation amendments
$\widetilde{\mathfrak{h}}$ remained smooth, straightforward and
compatible with the parallel refinements of the measurements of the
stationary bound-state energy spectra. An immanent limitation of
applicability of the traditional self-adjoint local-interaction
models emerged only after a next-step transition to the description
of the motion of a relativistic electron (or electrons) in some
perceivably stronger (e.g., heavy-ion) central Coulomb potential
$\widetilde{\mathfrak{v}} = -Ze^2/|\vec{r}|$. At the overcritical
effective charges with $Z> 1/\alpha \approx 137$, due to the so
called Klein paradox, the system crossed the boundary of stability
and entered a ``mathematically forbidden'' zone. The ground-state
energy ceased to be observable since it acquired a non-vanishing
imaginary part (cf., e.g., pp. 195 -- 206 in \cite{FlueggeII} for
details). Similar problems also occurred in the case of Klein-Gordon
equation (cf. \cite{Constantinescu}).

The loss-of-observability process of the degeneracy and of the
subsequent complexification of the energy levels can be interpreted
as a non-conservative quantum phase transition \cite{Denis,DenisII}.
In the language of physics the phenomenon is traditionally
attributed to the emergence of a new relevant degree of freedom
\cite{Landau,Nimrod}. In 1998, Bender with Boettcher \cite{BB}
proposed an alternative mathematical interpretation of the
phenomenon. They pointed out that within the conventional Hermitian
formulations of quantum theory the quantitative description of the
quantum phase transition phenomena is difficult, mainly because one
has to interrelate the unitary quantum world with the non-unitary
quantum world (cf. also the Jones' dedicated study
\cite{interface}).

The latter observation served as an immediate inspiration of our
present study. The paper will be organized as follows. In
introductory section \ref{prva}) we shall recall a few facts about
${\cal PT}-$symmetry and about quantum phase transitions. In section
\ref{tretisec} we then introduce, via a schematic model, the key
concept of our paper, viz, the notion of the so called
Hermitian--non-Hermitian interface. This material will be followed
by sections \ref{nuce} and \ref{ctvrtasec} in which we explain that
for our present purposes the stationary non-Hermitian
Schr\"{o}dinger-picture description of quantum dynamics of reviews
\cite{Geyer,Carl,ali} would not suffice, and that we shall need its
generalized, time-dependent and hiddenly Hermitian versions as
described in Refs.~\cite{timedep,SIGMA} and as critically reviewed,
more recently, in \cite{NIP}. A compact summary of the theory will
be provided, listing the dynamical equations in section \ref{nuce}
and then turning attention to some of the purely phenomenological
aspects of the resulting picture of physics in section
\ref{ctvrtasec}. In section \ref{patasec} we then return again from
the abstract theoretical lesson to our concrete illustrative
benchmark model. We use it to explain, in some technical detail,
some overall features of the process of the abstract construction of
the appropriate physical Hilbert space ${\cal H}^{(second)}$ and of
its concrete representation (based on the mere amendment of the
inner product) in the more friendly (albeit manifestly unphysical)
Hilbert space ${\cal H}^{(first)}$. The text will be then completed
by the two shorter sections of Discussion and Conclusions.

\section{${\cal PT}-$symmetry and its breakdown\label{prva}}

In the literature devoted to quantum systems and to the questions of
their stability theoreticians are usually clearly separating the
conventional Hermitian theories (in which the energies are assumed
real and in which the evolution is unitary) from the traditional
versions of non-Hermitian theories which deal, exclusively, with
unstable and resonant quantum systems. An explanation of such a
split of scope may be found in chapter 10 of monograph
\cite{Nimrod}. The author's attention is paid there to the latter,
complex-energy models. A parallel outline of the current
understanding of the unitary, stable quantum systems may be sought,
e.g., in the most recent collection of reviews \cite{book}. In our
present paper we shall restrict our attention just to the latter
subclass of the quantum models and phenomena in which the energies
remain real even if the representation of the observables themselves
becomes non-Hermitian, viz., quasi-Hermitian \cite{Geyer} or ${\cal
PT}-$symmetric \cite{Carl} or pseudo-Hermitian \cite{ali}.

\subsection{Bound states: The loss of observability}

The main purpose of our present paper is a clarification of several
paradoxes which were mentioned, in the literature, after the
publication of the Bender's and Boettcher's influential letter
\cite{BB}. The conventional Hermitian formulation of quantum
mechanics has been declared there over-restrictive. The authors
illustrated their idea via a hypothetical, manifestly non-Hermitian
ordinary differential Hamiltonian
 \be
 H=H^{(BB)}(\delta)=-\frac{d^2}{dx^2} +
 x^2({\rm i}x)^\delta \neq H^\dagger\,
 \label{selat}
 \ee
living, for $\delta \in (0,2)$ at least, in the entirely
conventional Hilbert space $L^2(\mathbb{R})$. They came to the
conclusion that in spite of the manifest non-Hermiticity of the
operator its energy spectrum is real, discrete and bounded from
below, i.e., potentially observable (rigorously, the conjecture has
been proved in \cite{DDT}). This property has been attributed to the
${\cal PT}-$symmetry of the Hamiltonian where ${\cal P}$ means
parity while the antilinear operator ${\cal T}$ simulates
time-reversal \cite{Carl}.

At the beginning of the new millennium the resolution of the
apparent contradiction between the manifest non-Hermiticity of the
Bender's and Boettcher's Hamiltonian and the strict reality of the
bound-state energies has been found in the older, quasi-Hermitian
formulation of quantum mechanics \cite{Geyer,ali}. The idea has been
made widely accepted, mostly under the name of ${\cal PT}-$symmetric
quantum mechanics (PTQM).

As a consequence of the subsequent developments, the modern
theoretical description of the unitary quantum systems is now
already partitioned into the so called Hermitian and non-Hermitian
branches. The former branch is thoroughly explained in conventional
textbooks \cite{Messiah}. For an introduction in the latter,
PTQM-inspired philosophy the reader is recommended to consult, e.g.,
the well-written reviews \cite{Carl,ali} (cf. also the
non-stationary extension of the formalism as introduced in
\cite{timedep} and reviewed in \cite{SIGMA}). It is worth
emphasizing that only the proper use of the PTQM-inspired
formulation of quantum theory endowed, e.g., the first-quantized
Klein-Gordon equation of textbooks, almost a full century after its
introduction, with a correct and consistent probabilistic
interpretation~\cite{aliKG}.

From the pragmatic point of view of experimental physics one of the
most important innovations characterizing the nonstandard PTQM
models may be seen in their capability of reaching the very boundary
of the unitary and stable dynamical regime. For example, the
spectrum of the toy model (\ref{selat}) is real (i.e., in principle,
compatible with the unitarity of the evolution) at all of the
non-negative exponents $\delta \geq 0$. This spectrum, nevertheless,
immediately loses these properties at an arbitrarily small negative
$\delta<0$ where the reality of the energies only survives for a
finite, $\delta-$dependent number of the low-lying levels~\cite{BB}.

In the extensive dedicated literature, the sudden loss of the
stability of the system at certain parameters and couplings is
usually interpreted as the spontaneous breakdown of ${\cal
PT}-$symmetry \cite{Carl}. This loss may be interpreted as a quantum
phase transition of the first kind \cite{Denis} and/or as a
cusp-like quantum catastrophe \cite{catast}. Various simulations of
such an abrupt loss of the observability of the energy made the PTQM
formalism also enormously popular among mathematicians \cite{book}
and non-quantum theoreticians and experimental physicists
\cite{Makris,Musslimani}.

\subsection{Scattering states: Giving up the unitarity}

During the birth of the PTQM formalism the conventional self-adjoint
phenomenological Hamiltonians $\mathfrak{h}=\mathfrak{h}^\dagger$
(with the robust reality of the spectrum)  were declared not
sufficiently flexible. The innovated models sampled by
Eq.~(\ref{selat}) were found compatible with the above-mentioned
hypothetical, phenomenologically well motivated requirement of the
possibility of a merger and of a subsequent complexification of the
energy eigenvalues. In the context of physics the phenomenon of such
a type has been interpreted as a spontaneous breakdown of the ${\cal
PT}-$symmetry of the system (see, e.g., the physics-oriented review
\cite{Carl} for details). The mathematical essence of the necessary
generalization of the formalism has been reconfirmed to lie in the
non-Hermiticity of the operators with real spectra. In the benchmark
model~(\ref{selat}) even the technical constraint $\delta \in (0,2)$
has been found removable. After an {\it ad hoc}, $\delta-$dependent
amendment of the integration contour of $x$ the spectrum has been
shown to stay real for all of the non-negative real exponents
$\delta \in [0,\infty)$.

The loss of the reality of the spectrum occurred at $\delta=0$. It
was not too difficult to conclude that the evolution controlled by
the toy-model Hamiltonian (\ref{selat}) is deeply non-unitary at
$\delta<0$. A more sophisticated interpretation of the evolution at
$\delta \in (0,2)$ or at $\delta \in (0,\infty)$ was needed of
course, but the necessary amendment of the theory has been developed
soon. In brief, the evolution has been found unitary in an {\it ad
hoc\,} Hilbert space ${\cal H}^{(physical)} \neq L^2(\mathbb{R})$ in
which the Hamiltonian itself (which is non-Hermitian in
$L^2(\mathbb{R})$ where we write $H \neq H^\dagger$) is
reinterpreted as self-adjoint \cite{ali}.

One of the limitations of the applicability of the PTQM approach was
revealed by Jones \cite{preinterface}. He turned attention to the
dynamical regime of unitary quantum scattering and replaced the
bound-state Hamiltonian (\ref{selat}) by the point-interaction toy
model
 \be
 H=H^{(HJ)}(\alpha,\beta,L)= -\frac{d^2}{dx^2} + \alpha V_{(H)}(x)
 + {\rm i} \beta V_{(N)}(x,L)\,.
 \label{znovusame}
 \ee
The conventional attractive delta-function interaction
$V_{(H)}(x)=-\delta(x)$ [which is Hermitian in $L^2(\mathbb{R})$]
was complemented there by a non-Hermitian but ${\cal PT}-$symmetric
partner $V_{(N)}(x,L)=\delta(x-L)-\delta(x+L)$. After a detailed
analysis of the model the Jones' ultimate recommendations were
discouraging. He came to the conclusion that one cannot accept the
fact that in the non-Hermitian picture the scattering in-state waves
``enter from both left and right'' and that there exists no
Hermitian/non-Hermitian-interaction interface, i.e., in his
interpretation, a {\em spatial\,} separation distance $L \approx
L_0$ at which one could treat model (\ref{znovusame}) as {\em
both\,} quasi-Hermitian {\em and\,} purely Hermitian. In his own
words, ``the physical picture changes drastically when going from
one picture to the other'' so that ``the only satisfactory
resolution of this dilemma is to treat the non-Hermitian scattering
potential as an effective one, and work in the standard framework of
quantum mechanics, accepting that this effective potential may well
involve the loss of unitarity when attention is restricted to the
quantum mechanical system itself and not its environment''
\cite{interface}.

\section{Phase transitions in a benchmark model\label{tretisec}}

In contrast to the abrupt and drastic physics-changing losses of
observability and/or of causality as mentioned above, the more
subtle and ``softer'', energy-conserving quantum phase transitions
of the second kind do not seem to have attracted the attention of
the experimentalists yet. We believe that such an attention could be
re-attracted by our forthcoming conceptual considerations.

A rarely emphasized theoretical possibility of the energy-conserving
phase-transition processes of the second kind has already been
noticed to exist in several less popular ${\cal PT}-$symmetric toy
models \cite{ptho,sdavidem,sdenisem}. Here, we are reopening the
discussion. We are persuaded that, in particular, the Jones'
scepticism is mathematically correct but that it is based, in the
context of physics, on a rather subtle misunderstanding. At an
arbitrary {\em fixed\,} set of parameters, indeed, the
quasi-Hermitian and Hermitian pictures of reality {\em must be}, by
definition, strictly equivalent \cite{ali}. In other words, the
concept of a Hermitian/non-Hermitian-interaction interface {\it
alias\,} Hermitian--quasi-Hermitian phase transition can only be
introduced as a specific, model-dependent set of parameters $\,{\cal
D}_{(inter\!face)}$ at which the Hermitian and quasi-Hermitian
representations of a quantum system would {\em
coincide}~\cite{ahep}.

\subsection{The existence of interface\label{nasmo}}

\begin{figure}[h]                    
\begin{center}                         
\epsfig{file=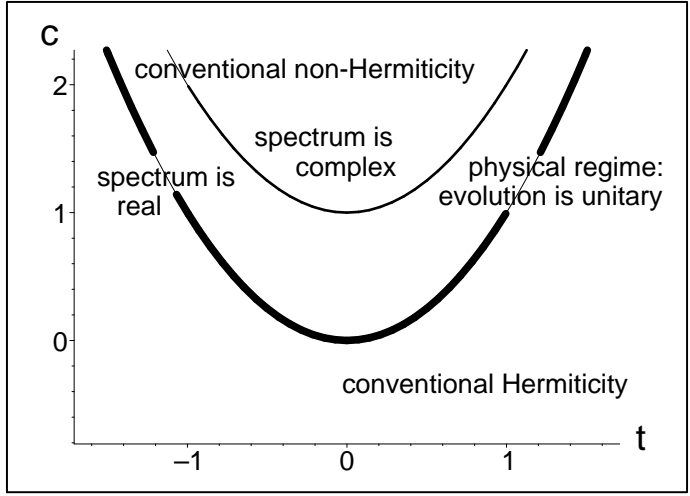,angle=270,width=0.56\textwidth}
\end{center}    
\vspace{2mm} \caption{The loss of the conventional Hermiticity with
the growth of $c$ across the thick curve, and the ultimate end of
the observability of the energy in $c-t$ plane (the spontaneous
breakdown of ${\cal PT}-$symmetry, thin curve) for toy-model
Hamiltonian~(\ref{primo}).
 \label{uba2a}
 }
\end{figure}

\begin{figure}[h]                    
\begin{center}                         
\epsfig{file=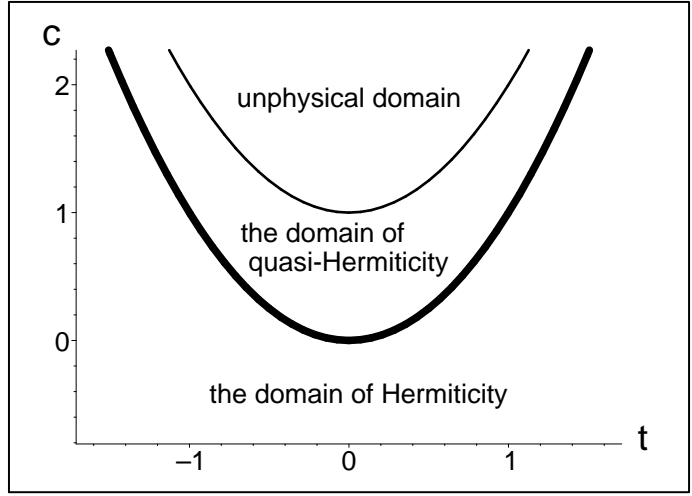,angle=270,width=0.56\textwidth}
\end{center}    
\vspace{2mm} \caption{Hermitian--quasi-Hermitian interface (thick
curve, Eq.~(\ref{interfa})).
 \label{uba2be}
 }
\end{figure}

According to the conventional, Hermitian quantum theory of textbooks
the parameter- and time-dependent family of matrices
 \be
 \mathfrak{h}(c,t) =
 \left[ \begin {array}{cc} -1&{\rm i} \sqrt {{t}^{2}-c}\\
 \noalign{\medskip}-{\rm i}\sqrt {{t}^{2}-c}&1\end {array} \right]
 \label{primo}
 \ee
can be perceived as an elementary sample of a phenomenological
Hamiltonian representing a stable quantum system ${\cal
S}_{(conventional)}$ if an only if the matrix is Hermitian in the
preselected physical Hilbert space, i.e., say, in
 $
 {\cal
 H}^{}_{(conventional)}=\mathbb{C}^2$,
 \be
 \mathfrak{h}(c,t) =\mathfrak{h}^\dagger(c,t)\,, \
  \ \ \ {\rm }\ \ \ \ \  t^2-c
 \geq 0.
 \label{seprimo}
 \ee
The evolution of the underlying quantum system will be unitary due
to the Stone theorem \cite{Stone}. In Schr\"{o}dinger picture this
evolution will be controlled by the conventional Schr\"{o}dinger
equation
 \be
 {\rm i}\frac{d}{dt} |\psi(t)\pkt = \mathfrak{h}(c,t)
 |\psi(t)\pkt\,,
 \ \ \ \ \ \
 |\psi(t)\pkt \ \in \  {\cal H}_{(conventional)}^{}\,.
 \label{SEq}
 \ee
In the plane of parameters $c$ and $t$ the set of admissible values
 $$
 {\cal D}_{(conventional)} = \{(c,t)\ | \ c \leq t^2\}
 $$
will fill the space on, and below, the lower, thicker parabola of
Figs.~\ref{uba2a} and~\ref{uba2be}.



%

According to the slightly less conventional versions of quantum
theory (cf. review \cite{ali}) the candidates (\ref{primo}) for
Hamiltonians may be made phenomenologically acceptable whenever the
energies remain real and non-degenerate, i.e., potentially
observable. Thus, after an {\it ad hoc\,} amendment of the physical
Hilbert space the conventional Hermiticity requirement can be
generalized and replaced by a more sophisticated but less
restrictive hidden Hermiticity called quasi-Hermiticity
\cite{Dieudonne}. For our matrix (\ref{primo}) such an innovation of
the theory would imply that the unitarity of the evolution of the
underlying quantum system can be guaranteed even in a non-Hermitian
dynamical regime or, more precisely, whenever the easily evaluated
eigenvalues
 \be
 E_\pm(c,t) = \pm \sqrt{t^2+1-c}\,
 \label{eneleve}
 \ee
satisfy the much weaker reality and non-degeneracy constraint.
Besides the above-mentioned ``Hermitian'' quantum systems ${\cal
S}_{(conventional)}$ one can, therefore, speak also about the
non-Hermitian but still unitary quantum systems ${\cal
S}_{(quasi-Hermitian)}$.

\subsection{Quasi-Hermitian regime}

In the light of Eq.~(\ref{eneleve}) the quasi-Hermitian extension of
the scope of quantum theory is feasible if and only if the
parameters $c$ and $t$ stay confined inside a complementary open set
of admissible parameters,
 \be
  {\cal D}_{(quasi-Hermitian)}= \{(c,t)\ | \ t^2 < c< t^2+1\}\,.
  \ee
In Figs.~\ref{uba2a} and~\ref{uba2be} such a ``quasi-Hermiticity''
domain of the new eligible parameters fills the curved-stripe space
between the two parabolas. The thicker parabolic curve lies in the
middle of the phenomenologically admissible physical domain
 $$
  {\cal D}_{(admissible)}=  {\cal D}_{(conventional)}\bigcup
   {\cal D}_{(quasi-Hermitian)}\,
 $$
which fills the whole space below the thinner parabola. Obviously,
the lower, thick curve represents a well defined  boundary between
the Hermitian (= lower) and quasi-Hermitian (= upper)
unitary-evolution regimes. The Hamiltonian itself is diagonal along
this curve. We will call this curve an ``interface'',
 \be
 {\cal D}_{(inter\!face)} = \{(c,t)\ | \ c = t^2\}\,.
 \label{interfa}
 \ee
We shall also slightly change here a few other notation conventions.
Firstly, our toy model Hamiltonian will {\em exclusively\,} be
written in the lower-case format of Eqs.~(\ref{primo}) +
(\ref{seprimo}) {\em on and below\,} the interface curve
(\ref{interfa}) of Figs.~\ref{uba2a} and~\ref{uba2be}. In parallel,
the {\em same\,} matrix will be {\em always\,} denoted by the
upper-case symbol whenever the parameters get chosen {\em above\,}
the interface, making the matrix non-Hermitian,
 \be
 H(c,t) =
 \left[ \begin {array}{cc} -1& \sqrt {c-{t}^{2}}\\
 \noalign{\medskip}-\sqrt {c-{t}^{2}}&1\end {array} \right]
 \ \neq \ H^\dagger(c,t)\,,
 \ \ \ \
 c>t^2
 \,.
 \label{rosecumo}
  \ee
The key purpose of such a restriction is to underline that the
physics behind the different symbols (viz., behind
$\mathfrak{h}(c,t)$ with $c<t^2$ and behind $H(c,t)$ with $c>t^2$)
is different. For two reasons. Not only because the parameters are
different but, first of all, because in contrast to $H(c,t) \neq
H^\dagger(c,t)$, the self-adjoint operator $\mathfrak{h}(c,t)$ may
be assigned the conventional spectral representation (cf. the
account of such an aspect of the theory in \cite{SKbook}).

Let us conclude that we do not need to care about the description of
the system in the Hermitian dynamical regime. This would be purely
routine and the details can be left to the reader. In contrast,
after the system passes the interface and enters the quasi-Hermitian
domain, multiple technical aspects of its description become far
from trivial.

\section{Non-stationary quasi-Hermitian dynamics \it in
nuce\label{nuce}}

We shall need the {\em time-dependent\,} extension of the stationary
PTQM formalism of Refs.~\cite{Carl,ali} in the form which has been
proposed in Ref.~\cite{timedep}. The name of ``three-Hilbert-space
(3HS) formulation of quantum mechanics'' was coined and advocated
for this upgrade of the theory in subsequent compact reviews
\cite{SIGMA,NIP,MZbook}. Nontrivial applications of the resulting
non-stationary 3HS approach are currently being sought
\cite{MZbook,IJTP,Wang,Maamache,FringMou}. Attention is being
shifted from the stationary context to the general time-dependent
scenario in which the quasi-Hermitian (i.e., in our notation,
upper-case) and time-dependent generic Hamiltonians $H(t)$ are
treated as isospectral to their Hermitian (i.e., in our notation,
lower-case) alternative representatives
 \be
  \mathfrak{h}(t)
 = \Omega(t)\,H(t)\,\Omega^{-1}(t)
 =\mathfrak{h}^\dagger(t)
 \,.
 \label{yelba}
 \ee
In our present study of matrix models, the {\it ad hoc\,}
construction of a suitable invertible (often called Dyson's) map
$\Omega(t)$ is just a routine linear-algebraic procedure. As long as
relation (\ref{yelba}) may be rearranged to read
 \be
 H^\dagger(t)\,\Theta(t) = \Theta(t)\,H(t)\,,
 \ \ \ \ \ \Theta(t)=\Omega^\dagger(t)\Omega(t)\,,
 \label{jedie}
 \ee
the matrix operator $H(t)$ may be declared quasi-Hermitian whenever
we manage to find $\Omega(t)$ and $\Theta(t)$ such that
Eq.~(\ref{jedie}) is satisfied.

\subsection{The doublet of Schr\"{o}dinger equations}

In Refs.~\cite{SIGMA,NIP} the ultimate, non-stationary 3HS version
of the quasi-Hermitian quantum theory is characterized as a
representation of a quantum system ${\cal S}_{(quasi-Hermitian)}$
which is based on the simultaneous use of the three representative
Hilbert spaces ${\cal H}^{(first)}$, ${\cal H}^{(second)}$ and
${\cal H}^{(conventional)}$. The ket-vectors $|\psi\kt \in {\cal
H}^{(first)}$ are assumed to coincide with the kets $|\psi\kt \in
{\cal H}^{(second)}$. In the nuclear-physics
exemplification~\cite{Geyer} both of them describe the ``effective''
bosons while the ``real'' nucleons, fermions,  have to be
represented by {\em different}, spiked-ket symbols $|\psi\pkt \in
{\cal H}^{(conventional)}$.

The mutual correspondence
 \be
 |\psi(t)\pkt = \Omega_{}(t)\,
 |\psi^{}_{}(t)\kt\,.
 \label{anza}
 \ee
between the kets is just a time-dependent generalization of the
Dyson's old idea \cite{timedep,Dyson}. In its spirit one inserts
ansatz (\ref{anza}) in the conventional lower-case Schr\"{o}dinger
equation sampled by Eq.~(\ref{SEq}) which lives in the Hilbert space
${\cal H}^{(conventional)}$ of textbooks. This insertion leads to
the equivalent equation
 \be
 {\rm i}\frac{d}{dt}\,|\psi(t)\kt=G(t)\, |\psi(t)\kt\,
   \label{bespeq}
 \label{SEtaujoa}
 \ee
which is defined in both of the spaces ${\cal H}^{(first)}$ and
${\cal H}^{(second)}$. We must only add that
 \be
 G(t) = H(t) - \Sigma(t)\,,
 \ \ \ \ \
 {\Sigma}_{}(t)
 = {\rm i}\Omega^{-1}_{}(t)\left [\frac{d}{dt}\,\Omega_{}(t)\right ]
  \,.
 \label{generato}
 \ee
We have to remind the readers that our instantaneous-energy-operator
$H(t)$ is assumed to be defined in a ``friendly'' Hilbert space
${\cal H}^{(first)}$ in which it is non-self-adjoint even though it
possesses the real (i.e., in principle, observable) spectrum. This
is the reason why another, ``sophisticated'' Hilbert space ${\cal
H}^{(second)}$ had to be introduced:

\begin{itemize}

\item
the space ${\cal H}^{(second)}$ is physical -- it is constructed as
unitarily equivalent to ${\cal H}^{(conventional)}$, i.e., to the
Hilbert space of textbooks;

\item
the conventional space is, by assumption, ``prohibitively
complicated'' and useless \cite{Geyer,Dyson}. Simplifications are
expected from our working in ${\cal H}^{(second)}$ \cite{ali};

\item
the manifestly unphysical Hilbert space ${\cal H}^{(first)}$ is
assumed friendly. It is, therefore, used as a carrier of the
representation of ${\cal H}^{(second)}$ (realized via a modification
of the inner product).

\end{itemize}

 \noindent
Conceptually, the latter representation is easy. It merely requires
a replacement of the antilinear Hermitian conjugation defining the
first space,
 \be
 {\cal T}^{(first)}: |\psi\kt \to \br \psi|\ \in \ \left
 ( {\cal H}^{(first)}
 \right )'
 \ee
by its second-space bra-to-curlyket analogue
 \be
 {\cal T}^{(second)}: |\psi\kt \to \bbbr \psi|\ \in \ \left ( {\cal
 H}^{(second)}
 \right )'\,.
 \ee
For this purpose it is sufficient to postulate the identification
 \be
 |\psi_{}(t)\pkt=\Omega_{}(t)|\psi_{}(t)\kt
 =\left [\Omega_{}^\dagger(t)\right ]^{-1}|\psi_{}(t)\kkkt
 \,.
  \label{CSPouSE}
 \ee
One of the most immediate consequences is the validity of the
conjugate-evolution Schr\"{o}dinger equation
 \be
 {\rm i}\frac{d}{dt}\, |{\psi}_{}(t)\kkkt =
  G_{}^\dagger(t)\,|{\psi}_{}(t)\kkkt\,.
 \label{SEtaujobe}
 \ee
Another, equally useful consequence is that the conventional
textbook orthogonality and/or orthonormality relations in the
``prohibited'' space ${\cal H}^{(conventional)}$ become equivalent
to the biorthogonality and/or biorthonormality relations in the
``recommended'' space ${\cal H}^{(second)}$,
 \be
 \pbr \psi_1|\psi_2\pkt=
 \bbbr \psi_1|\psi_2\kt\,.
 \ee
Once the latter space is always represented in the friendly,
auxiliary Hilbert space ${\cal H}^{(first)}$, one simply defines
$|\psi\kkkt = \Theta|\psi\kt$ \cite{SIGMA}.

\subsection{The doublets of Heisenberg equations \label{heitype}}

The variability of the Dyson maps with time forces us to realize
that even if the observable quantity represented by a lower-case
operator $\lambda$ is chosen time-independent, its isospectral
partner would vary with time, anyhow. Still, in \cite{NHeisenberg}
we revealed that the work with the non-stationary lower-case
operators is truly tedious so that we will keep assuming here that
$\lambda=\lambda^\dagger\neq \lambda(t)$ in ${\cal
H}^{(conventional)}$. Then we may Dyson-map
 \be
 \lambda \ \to \ \Lambda(t)=\Omega^{-1}(t)\,\lambda\ \Omega(t)\,.
 \label{defia}
 \ee
The straightforward differentiation of this product leads to the
operator differential equation of Heisenberg type,
 \be
 {\rm i}\frac{d}{dt}\, {\Lambda}_{{}}(t)
  = {\Lambda}_{{}}(t) {\Sigma}_{{}}(t)
 -{\Sigma}_{{}}(t) {\Lambda}_{{}}(t)
 \,.
 \label{fdeveta}
 \ee
Again, the self-adjointness in ${\cal H}^{(conventional)}$ is
translated into the quasi-Hermiticity in ${\cal H}^{(first)}$,
 \be
 {\Lambda}_{{}}^\dagger (t)= \Theta(t)\Lambda(t)\,\Theta^{-1}(t)\,.
 \label{mutual}
 \ee
Whenever we wish to keep the trace of the observability explicit, it
makes sense to work, in parallel, with the second Heisenberg
equation
 \be
 {\rm i}\frac{d}{dt}\, {\Lambda}^\dagger_{{}}(t)
  = {\Lambda}_{{}}^\dagger(t) {\Sigma}_{{}}^\dagger(t)
 -{\Sigma}_{{}}^\dagger(t) {\Lambda}_{{}}^\dagger(t)
 \,.
 \label{fdevetbe}
 \ee
In the special case of the observable energy $H(t)$ in the
conservative scenario with $\mathfrak{h} \neq \mathfrak{h}(t)$ the
pair of Eqs.~(\ref{fdeveta}) and (\ref{fdevetbe}) could be also used
and, in this case, modified,
 \be
 {\rm i}\frac{d}{dt}\, {H}_{{}}(t) =
 {G}_{{}}(t) {H}_{{}}(t)-{H}_{{}}(t) {G}_{{}}(t)
 \,.
 \label{hadeveta}
 \ee
An independent comment can be made concerning the metric operator
for which one starts from the elementary mathematical identity
 \be
 {\rm i}\frac{d}{dt}\, {\Theta}_{{}}(t) =
 {\Theta}_{{}}(t) {\Sigma}_{{}}(t)-{\Sigma^\dagger}_{{}}(t)
 {\Theta}_{{}}(t)
 \ee
and deduces its equivalent alternative
 \be
 {\rm i}\frac{d}{dt}\, {\Theta}_{{}}(t) =
 {G^\dagger}_{{}}(t)
 {\Theta}_{{}}(t)-{\Theta}_{{}}(t) {G}_{{}}(t)\,.
 \label{26}
 \ee
In recent paper~\cite{NIP} the direct solution of Eq.~(\ref{26}) was
discouraged as tedious, inefficient and not always necessary. Still,
one could try to solve this differential equation for operator
$\Theta(t)$, numerically, in some sufficiently elementary special
cases. This was done, e.g., by Hynek B\'{\i}la \cite{Bila} and,
later, by several other teams of researchers \cite{FringMou}. All of
these authors revealed and pointed out that the resulting operators
of the Hilbert space metric $\Theta(t)$ seem to be enormously
sensitive not only to the properties of the generators $G(t)$ but
also to the initial choice of $\Theta(t)$ at the preparation time,
i.e., in our present physical context, at the interface, i.e., at
the instant $t=t_0$ of the phase transition.

\section{Physics of quasi-Hermitian Hamiltonians  \it in
nuce\label{ctvrtasec}}

In our present model-based analysis of the phenomena connected with
the existence of the Hermitian--quasi-Hermitian interface we may
follow the conventional textbooks and use trivial $ \Omega(c,t)=I$
whenever $(c,t)\in {\cal D}_{(conventional)}$, i.e., in the
Hermitian regime. In the more sophisticated regime with $(c,t)\in
{\cal D}_{(quasi-Hermitian)}$, a non-trivial $ \Omega(c,t)\neq I$
will be needed. After such a generalization the formalism becomes
perceivably more complicated.

\subsection{Terminology}

The information about the (say, unitary) time-evolution of a given
quantum system ${\cal S}$ can be carried by its wave function
$\psi(t)$ (one then speaks about the Schr\"{o}dinger picture of the
reality, SP \cite{Schrpic}), or by the relevant observables
${\mathfrak{q}}(t)$ (in the so called Heisenberg picture, HP
\cite{Heispic}), or both (in the universal Dirac's {\it alias\,}
interaction picture, IP \cite{Messiah}). In the light of the recent
theoretical developments (cf., e.g., dedicated book \cite{book}) one
can further distinguish between the so called Hermitian and
non-Hermitian versions of the respective alternative formulations of
the quantum dynamical laws. Thus, the traditional reviews of the
Hermitian formulations (e.g., \cite{Styer}) may be complemented by
the detailed outlines of the non-Hermitian Schr\"{o}dinger picture
(NSP, \cite{Carl,ali}), of the non-Hermitian Heisenberg picture
(NHP, \cite{NHeisenberg,Lee}) and/or of the non-Hermitian
interaction picture (NIP, \cite{timedep,SIGMA,NIP,Fring}).

The shared feature of all of the innovative non-Hermitian pictures
is that they work with the operators of observables (say, $Q(t)$)
which are only non-Hermitian in an auxiliary, mathematically
strongly preferable and technically friendly but plainly unphysical
Hilbert space ${\cal H}^{(first)}$. In this sense the widespread use
of the term ``non-Hermitian operators'' (so that one writes
$Q(t)\neq Q^\dagger(t)$) is mathematically correct (in ${\cal
H}^{(first)}$) but conceptually misleading. This is because all of
our ``exotic'' observables $Q(t)$ may be reinterpreted as
traditional and self-adjoint immediately after one replaces the
auxiliary, ``false'' Hilbert space ${\cal H}^{(first)}$ by its
manifestly physics-representing ``standard'' amendment ${\cal
H}^{(second)}$.

The survival of the misleading terminology had a few pragmatic
and/or historical reasons. The main one is that the ``correct''
physical Hilbert space ${\cal H}^{(second)}$ is in fact never used
in practice. In the majority of applications it is either
represented in ${\cal H}^{(first)}$ (see the preceding section) or
replaced by its unitarily equivalent avatar ${\cal
H}^{(conventional)}$. In the former case one should better write,
e.g., $Q(t)= Q^\ddagger(t):=\Theta^{-1}(t) Q^\dagger(t) \Theta(t)$.

In the literature, the notation is far from being unified. For
example, in the stationary quasi-Hermitian formalisms, our present
symbol $\Theta$ (denoting the special, time-independent physical
Hilbert-space metric) is just a Greek translation of the symbol $T$
used in the oldest review \cite{Geyer}. For the same operator, an
entirely different, subscripted symbol $\eta_+$ was introduced by
Mostafazadeh \cite{ali}. Equivalently, people also use the special
$\Theta$s  equal to the products ${\cal PC}$ of parity with the
(Hamiltonian-dependent) charge \cite{Carl}.

\subsection{Measurements}

From the point of view of experimentalists,  the evolution of any
quantum system controlled by the equations of preceding section must
be initiated by the preparation of the system (say, in a pure state)
at an ``initial'' time (say, at $t=0$). Subsequently, the
verification of the predictions is to be performed using the
measurement over the system at a suitable ``final'' time $t=T>0$. It
is well known that ``before the phase transition'', i.e., in the
conventional Hermitian regime the matrices
$\Omega(c,t)=\Theta(c,t)=I$ may be kept trivial so that, in the 3HS
language, all three Hilbert spaces coincide and the upper- and
lower-case Hamiltonians are allowed to coincide as well,
$\mathfrak{h}(c,t) = H(c,t)$. This means that in the conventional
dynamical regime one just follows the textbooks. The predictions
concerning, say, a time-dependent observable $\mathfrak{q}(t)$ are
simply obtained via the routine evaluation of the mean-value
overlaps
 \be
 \pbr \psi(T)|\mathfrak{q}(T)|\psi(T)\pkt\,
 \ee
where the time-dependence of the operator $\mathfrak{q}(t)$ is
assumed prescribed in advance and where the time-dependence of the
wave functions $|\psi(t)\pkt$ is obtained by the solution of
Schr\"{o}dinger Eq.~(\ref{SEq}).

After the quantum system in question passes the
Hermitian--quasi-Hermitian interface and after it continues evolving
in its quasi-Hermitian phase, the latter formula defined in ${\cal
H}_{(conventional)}$ may be declared intractable because we are only
given now our Hamiltonian (i.e., its toy model sample $H(c,t)$) in
its non-Hermitan form (cf. Eq.~(\ref{rosecumo})). Our overall
methodical assumptions force us to use the general 3HS formalism
and, in particular, the nontrivial forms of the Dyson maps
$\Omega(t)\neq I$. Also just the upper-cases representations of the
observables may be assumed tractable. In the light of
Eqs.~(\ref{yelba}) and (\ref{defia}) (in its specification
$Q(t)=\Omega^{-1}(t)\,\mathfrak{q}(t)\ \Omega(t)$) this means that
the lower-case representatives of the observables become unknown and
different from their upper-case avatars. Fortunately, in the general
3HS setting the knowledge the lower-case observables is superfluous
due to the identity
 \be
 \pbr \psi(T)|\mathfrak{q}(T)|\psi(T)\pkt
 =
 \bbbr \psi(T)|Q(T)|\psi(T)\kt\,.
 \label{simmy}
 \ee
Thus, it is sufficient to evaluate just the right-hand-side
expression in practice.

\subsection{Instantaneous energies}

Via Figs.~\ref{uba2a} and~\ref{uba2be} we discussed, in section
\ref{tretisec}, the influence and the consequences of the growth of
parameter $c$ at a given time $t$. For the purposes of the study of
dynamics in the language of the above-outlined non-stationary 3HS
formalism it makes better sense to keep the parameter $c$ unchanged
and to check what is happening during the evolution of the system in
time.

\begin{figure}[h]                    
\begin{center}                         
\epsfig{file=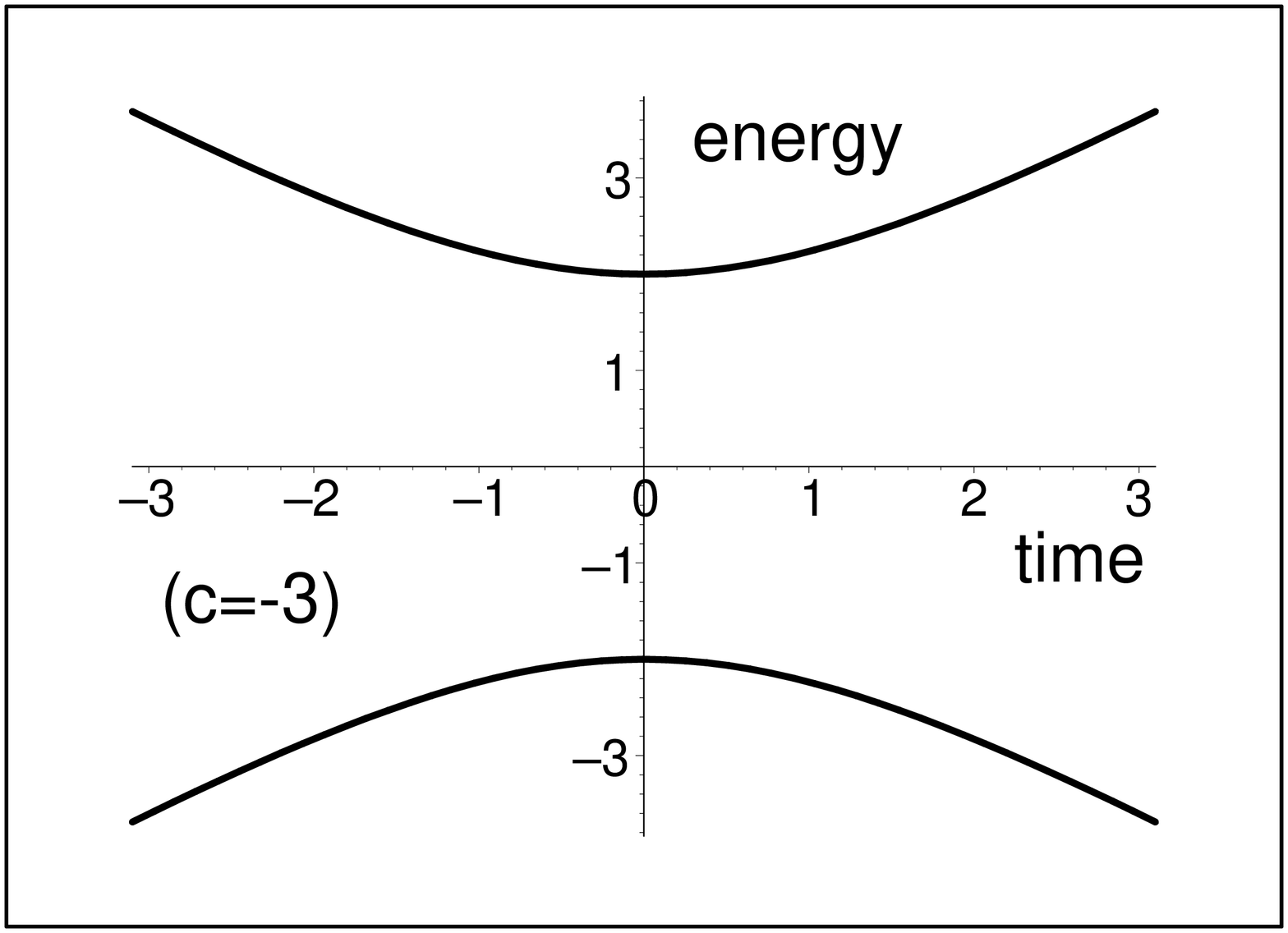,angle=270,width=0.36\textwidth}
\epsfig{file=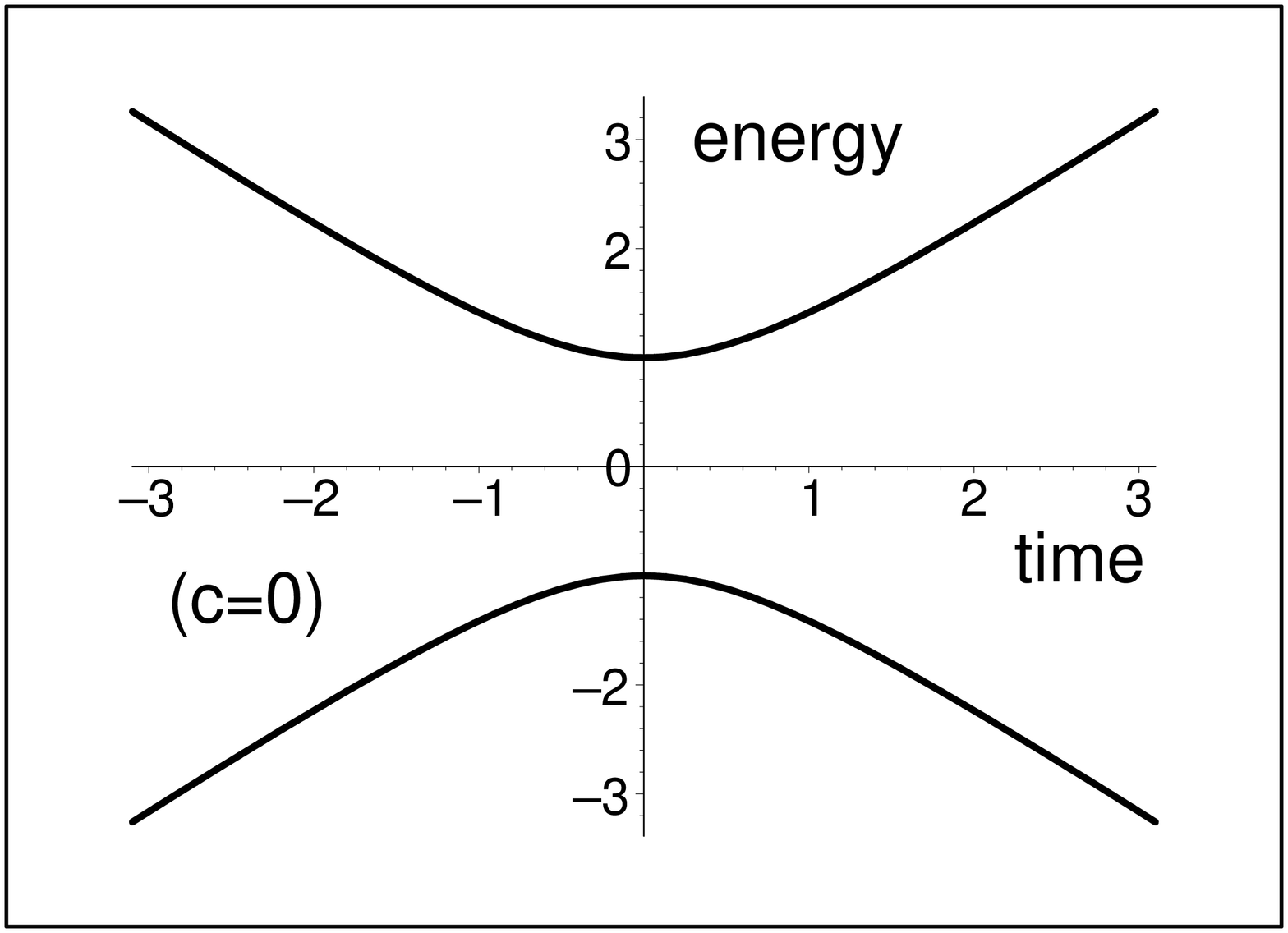,angle=270,width=0.36\textwidth}
\end{center}    
\vspace{2mm} \caption{Constant$-c$ energies $E_\pm(c,t)$ of
Eq.~(\ref{eneleve}) in Hermitian regime (left picture) and on its
boundary (right picture).
 \label{rija21}
 }
\end{figure}

What is of primary interest is the time-dependence of the (in
principle, measurable) instantaneous energies as prescribed by
Eq.~(\ref{eneleve}). In the two pictures of Fig.~\ref{rija21} let us
sample the conventional scenario in which the energies exhibit a
characteristic pattern of the so called ``avoided crossing''.

\begin{figure}[h]                    
\begin{center}                         
\epsfig{file=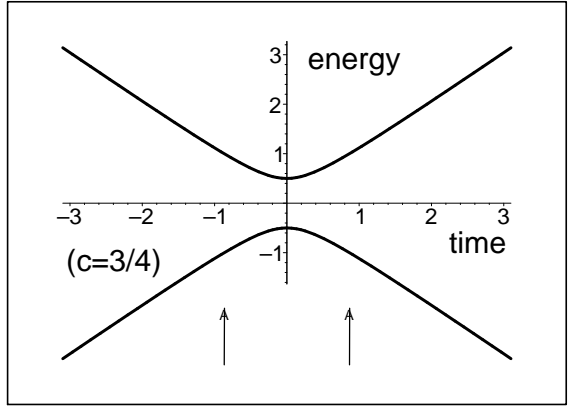,angle=270,width=0.46\textwidth}
\end{center}    
\vspace{2mm} \caption{Constant$-c$ energies~(\ref{eneleve}) in the
partially quasi-Hermitian regime.
The two arrows mark the
interval of non-Hermiticity
with boundaries $t_{(inter\!face)}=\pm \sqrt{c}\approx \pm 0.866$.
 \label{rija3}
 }
\end{figure}

In the subsequent example of Fig.~\ref{rija3} we see that the same
avoided crossing behavior remains unchanged even if we move to a
non-Hermitian but still quasi-Hermitian dynamical regime with $c \in
(0,1)$. From the spectrum itself one could hardly guess that a
nontrivial Hilbert-space metric $\Theta(t)\neq I$ must be
constructed in the interval of $t \in (-\sqrt{c},\sqrt{c})$. Inside
this interval we have $(c,t) \in {\cal D}_{(quasi-Hermitian)}$ so
that the probabilistic interpretation of the non-stationary quantum
system in question becomes nontrivial. The predictions of the
results of instantaneous measurements must be calculated using the
right-hand-side expression in formula (\ref{simmy}) of course. In
other words, besides the necessity of the solution of the pair of
Schr\"{o}dinger Eqs.~(\ref{SEtaujoa}) and (\ref{SEtaujobe}), also
the time-dependence of the generic observables must be deduced from
the solution of the underlying Heisenberg-like equations as
discussed in Ref.~\cite{NIP} in full detail, and as sampled in
subsection \ref{heitype} above.

\begin{figure}[h]                    
\begin{center}                         
\epsfig{file=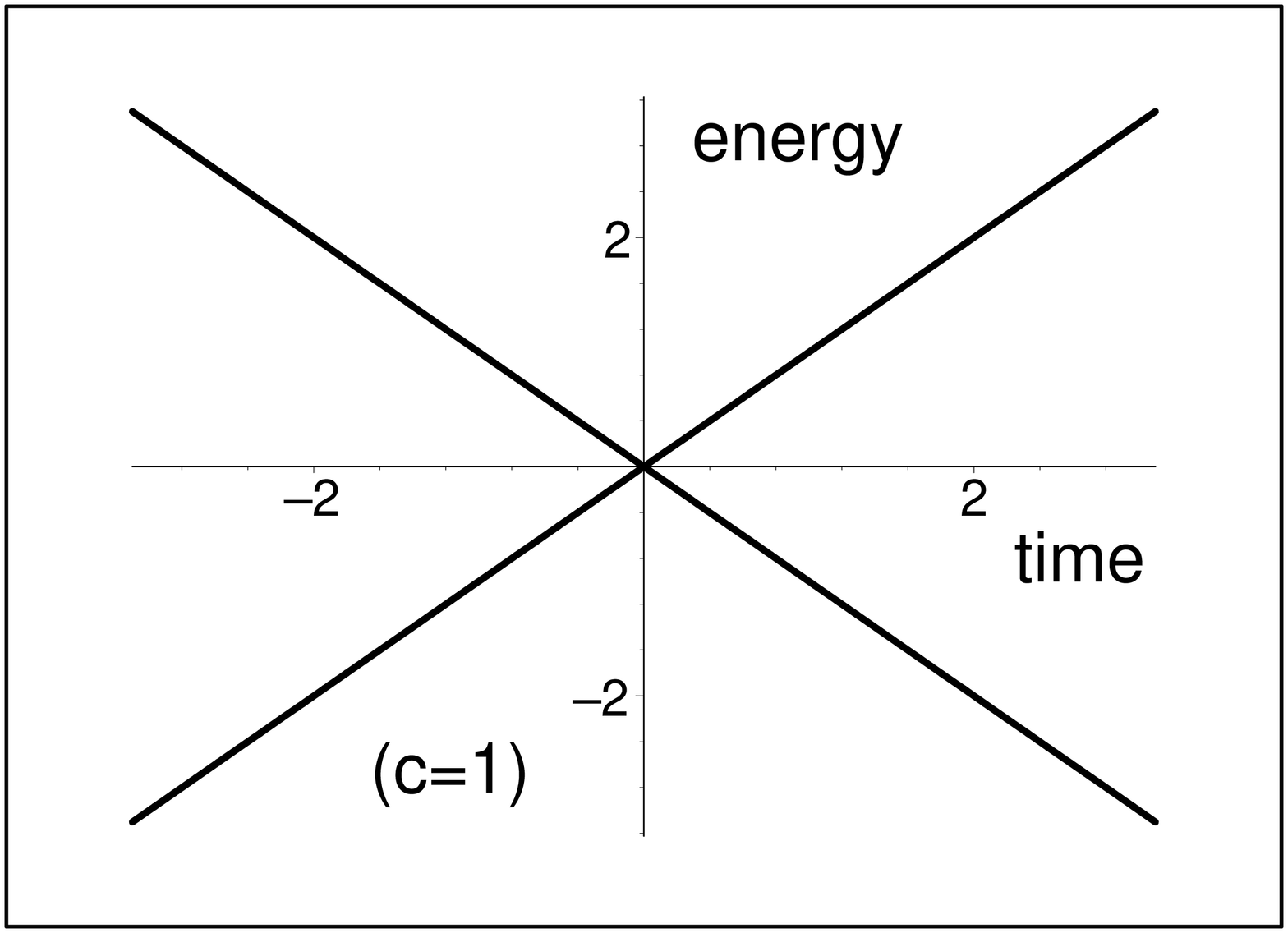,angle=270,width=0.36\textwidth}
\epsfig{file=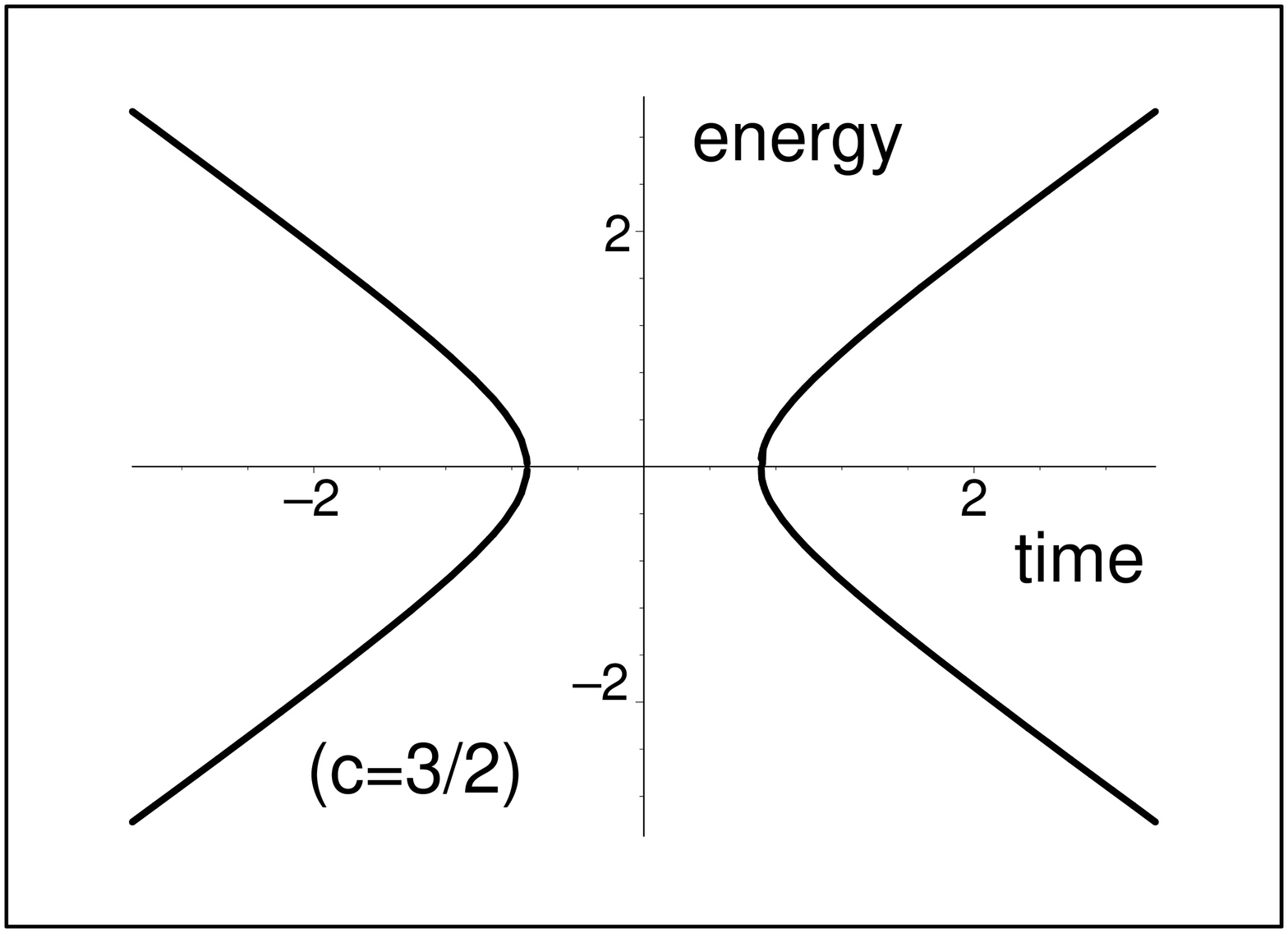,angle=270,width=0.36\textwidth}
\end{center}    
\vspace{2mm} \caption{Real parts of the constant$-c$
energies~(\ref{eneleve}) at the boundary of the quasi-Hermitian
regime (left picture: notice the degeneracy at $t=0$) and beyond
(right picture).
 \label{rija2}
 }
\end{figure}

We saw that in our toy models of paragraph \ref{nasmo} the
avoided-crossing behavior of the spectrum is shared by the
negative-parameter regime with $c<0$ (where the Hamiltonian is
safely Hermitian) and by the partially quasi-Hermitian regime at the
small and positive $c\in (0,1)$. The purely Hermitian description
only survives there at such times that $t^2>c$, cf.
Fig.~\ref{uba2a}. In contrast, a drastic qualitative change of the
spectrum occurs when we choose a larger value of $c\geq 1$. This is
illustrated by Fig.~\ref{rija2} where the left picture with $c=1$
samples the rather remarkable ``unavoided-crossing'' spectral
anomaly, and where the right picture samples the $c>1$ scenario in
which the energies cease to be real in the interval of $t \in
(-\sqrt{c-1},\sqrt{c-1})$.

The two instants $t_\pm = \pm \sqrt{c-1}$ of the ``quantum
catastrophe'' \cite{catast} mark the collapse of the system and
reflect the spontaneous breakdown of its ${\cal PT}-$symmetry
\cite{Carl}. In mathematics these values are called ``exceptional
points'' \cite{Kato}. Inside the interval of $t \in
(-\sqrt{c-1},\sqrt{c-1})$ our toy model Hamiltonian $H(c,t)$ of
Eq.~(\ref{rosecumo}) is not even quasi-Hermitian. It ceases to
describe any physical reality whatsoever. In contrast, once we
restrict attention, say, to the left half-line of time $t \in
(-\infty,-\sqrt{c-1}$, the importance of the
Hermitian--to--non-Hermitian interface (\ref{interfa}) is enhanced
because its existence now represents a gate and one of conditions of
the realization of the evolution leading to an ultimate fall of the
quantum system into instability. This makes the models with $c>1$
theoretically relevant, phenomenologically appealing and
methodically truly interesting.

\section{Benchmark model: Probabilistic interpretation
\label{patasec}}

Whenever we are given the operators of observables and whenever we
find the metric $\Theta$ compatible with relations (\ref{mutual}),
we may factorize $\Theta \to \Omega$ and formulate the dynamical
evolution equations. The recipe (cf. Ref.~\cite{NIP}) has thoroughly
been described above. For its present new application let us now
return to matrix (\ref{rosecumo}) with parameters $c$ (coupling)
and/or $t$ (time) localized, in the $c-t$ plane, between the two
parabolic curves of Fig.~\ref{uba2a}. Once we assume that the
parameters lie not too far from the lower parabola (i.e., from the
interface of our current interest), matrix (\ref{rosecumo}) can be
perceived as the operator of an observable energy. It characterizes
our hypothetical quantum system which was initially Hermitian but
which suffered the phase transition. This means that the parameters
were slightly changed. The system passed the
Hermitian--quasi-Hermitian interface but in the new dynamical regime
the evolution is still unitary.

\subsection{Metric operator}

In the quasi-Hermitian regime the physical contents of our real
Hamiltonian matrix $H(c,t)$ of Eq.~(\ref{rosecumo}) is given by the
real and symmetric matrix
 \be
 \Theta(c,t)=
 \left[ \begin {array}{cc} a(c,t)&b(c,t) \\
 \noalign{\medskip}b(c,t)&d(c,t)\end {array} \right]
 \ =\  \Theta^\dagger(c,t) >0
  \,
  \label{29}
 \ee
of the Hilbert-space metric. Compatibility condition (\ref{jedie})
may be checked to hold if and only if $b(c,t)=\gamma(c,t)u(c,t)$
where $\gamma(c,t)=\sqrt{c-t^2}$ and $u(c,t)=[a(c,t)+d(c,t)]/2$. In
this notation the energy eigenvalues (\ref{eneleve}) acquire the
transparent form $E_\pm = \sqrt{1-\gamma^2(c,t)}$ so that with
$\gamma(c,t)=\sin \tau(c,t)$ the whole information about our
quasi-Hermitian input Hamiltonian is reduced to the specification of
$\tau(c,t) \in (0,\pi/2)$. Effectively, this parameter measures the
distance from the interface (\ref{interfa}).

What remains for us to guarantee is the positivity of the metric.
This means that with $v(c,t)=a(c,t)d(c,t)$
both of its eigenvalues
 $
 \theta_\pm = u \pm \sqrt{(1+\gamma^2)u^2-v}>0
 $
must remain real (i.e., we must have $v\leq (1+\gamma^2)u^2$) and
positive (i.e., we must require $u>0$ and $v>\gamma^2u^2 $). This
enables us to reparametrize $v =v(u,\xi)=\gamma^2u^2+ u^2 \cos^2\xi$
with $\xi \in (0,\pi/2)$. The change $v \to \xi$ also simplifies
 $
 \theta_\pm = (1 \pm \sin \xi)\,u
 $.
Thus, the entirely general form of the metric will vary with the two
free parameters, viz., with $u=u(c,t) \in (0,\infty)$ and with
$\xi=\xi(c,t) \in (0,\pi/2)$.

The backward changes of parameters yielding the explicit form of
metric (\ref{29}) are trivial: The derivation of the formulae is
left to the readers. With this being done, the first step of the
construction of the model would be completed. What would have to
follow in applications would be the factorization of the metric into
Dyson maps $\Omega(c,t)$, the construction of operators
$\Sigma(c,t)$ and $G(c,t)$ (cf. Eq.~(\ref{generato})) and, finally,
the solution of Schr\"{o}dinger and Heisenberg equations.

\subsection{Physics near the interface}

One of the key messages delivered by the preceding subsection is
that due to the non-stationarity of our toy model Hamiltonian
$H=H(c,t) \equiv H[\gamma]$ one can select its physical
interpretation out of a two-parametric menu of the eligible metrics
$\Theta=\Theta(\gamma,u,\xi)$, i.e., of the physical Hilbert spaces
${\cal H}^{(second)}(\gamma,u,\xi)$, i.e., of the metric-dependent
sets of the quasi-Hermitian operators of observables
$\Lambda=\Lambda_j(\gamma,u,\xi)$ with $j=1,2,\ldots$ [cf.
Eq.~(\ref{mutual})].

The concrete specification of our present quasi-Hermitian model as a
system which was created by its passage through the interface has
two important methodical consequences. Firstly, it makes sense to
simplify our task and to restrict the scope of our analysis to the
very small vicinity of the interface, i.e., to the very small
(though still positive) values of the dynamical input-representing
parameter $\gamma \ll 1$, i.e., of the re-scaled time initiated as
the instant of the phase transition. Secondly, we have to postulate
that all of the changes of the system with time should be smooth. In
particular, this means that the two eigenvalues $\theta_\pm = (1 \pm
\sin \xi)\,u$ of the metric should be smooth functions of $\gamma$,
i.e., the difference $u-1$ and the size of the second parameter
$\xi$ should remain small at the small ``re-scaled times''
$\gamma\ll 1$.

Due to the elementary two-by-two matrix nature of our present
benchmark example we can make use of the available explicit formulae
and we could easily deduce the explicit forms of the corresponding
illustrative power-series expansions. It is methodically more
important to notice that the explicit construction of the
approximations can proceed, in fact, in an entirely
model-independent manner. One only has to consider a generic
quasi-Hermitian $\gamma-$dependent Hamiltonian (i.e., say, its
arbitrary non-Hermitian $N$ by $N$ real-matrix exemplification with
real eigenvalues) which is defined, near the interface, by its
Taylor series,
 \be
 H[\gamma]=H[0] + \gamma\,H'[0] + {\cal O}(\gamma^2)\,.
 \ee
On the interface we have $H[0]=H^\dagger[0]$ of course. We can
combine this general dynamical input information with a parallel
perturbation-series ansatz for the related metric near the
interface,
 \be
 \Theta[\gamma] = I + \gamma\,K + {\cal O}(\gamma^2)\,.
 \ee
After insertion in the Dieudonn\'{e}'s compatibility condition
(\ref{jedie}) this will yield the first-order perturbation version
of the constraint,
 \be
 H^\dagger[0]\,K-K\,H[0]=
 H'[0] -\left ( H'[0] \right )^\dagger\,.
 \ee
Routinely, this equation is to be solved, for $K$, by a suitable
linear algebraic algorithm. In similar spirit one could also proceed
in the higher-order perturbation constructions (advised by the
referee we relocated these technicalities to a more mathematically
oriented future publication).

\section{Discussion}

\subsection{Ambiguities}

In the present application of the non-stationary quasi-Hermitian
theory we were only given the observable of energy in its manifestly
non-Hermitian matrix representation~(\ref{rosecumo}). This implies
that the Dieudonn\'{e}'s Eq.~(\ref{mutual}) can only be interpreted
as a mere self-consistent restriction upon our choice of operators
(say, of $\mathfrak{h}(c,t)$ and $\Omega(c,t)$) rather than as their
unambiguous specification. This type of ambiguity was thoroughly
discussed in Ref.~\cite{Geyer}. In the context of physics the most
elementary method of the removal of the ambiguity of the
specification of operators $\mathfrak{h}(c,t)$ and $\Omega(c,t)$ may
be based on certain additional phenomenological assumption. Besides
the observability of the energy we may also require the existence of
another (generic) observable represented, say, by a self-adjoint
operator $\mathfrak{q}(c,t)=\mathfrak{q}^\dagger(c,t)$  [or
operators, not necessarily $(c,t)-$dependent] or by its/their
upper-case isospectral quasi-Hermitian avatar(s)
 \be
 Q(c,t)=\Omega^{-1}(c,t)\,\mathfrak{q}(c,t)\, \Omega(c,t)\,.
 \label{redysso}
 \ee
The mapping $\Omega(c,t)$ itself remains the same as before. This
means that the Hermiticities of the lower-case operators can be
simply reinterpreted as the respective Dieudonn\'{e}'s
\cite{Dieudonne} quasi-Hermiticity properties~(\ref{mutual}).

\subsection{Interfaces}

Quantum phase transitions are usually interpreted as a breakdown of
the unitarity of the evolution which is connected, in the light of
the well known Stone's theorem \cite{Stone}, with an abrupt change
of the effective Hamiltonian, i.e., with the sudden emergence of
some new relevant degrees of freedom. Still, there exist the quantum
evolution processes during which the Hamiltonian remains unchanged
and during which the responsibility for the phenomenon of the phase
transition is transferred to a redefinition of the underlying
physical Hilbert space. In 1992, for example, Scholtz et al
\cite{Geyer} introduced a sophisticated non-Hermitian boson-field
generalization of the so called Lipkin-Meshkov-Glick model. These
authors demonstrated (cf. Figure Nr. 1 in {\it loc. cit.}) that the
system exhibits a phase transition which is {\em not\,} caused by a
modification of the operators of the observables themselves.

In the real world of experimental physics the passage of a given
quantum system through its phase transition instant $t_0$ is usually
assumed to proceed very slowly, in an adiabatic dynamical regime.
Unfortunately, the authors of Ref.~\cite{Milburn} demonstrated that
in the non-Hermitian cases such an approximation strategy need not
be applicable in the quasi-Hermitian quantum mechanics in general.
In the light of the relevant review papers (cf., e.g., \cite{NIP}),
the transition from the well known Hermitian formalism of textbooks
to the slightly counterintuitive quasi-Hermitian picture of dynamics
may also lead to several other theoretical as well as purely
mathematical consequences. On the theoretical side one must
emphasize that after the passage through the interface the
observability status of the energies themselves remained, by the
construction of our illustrative model, unchanged. Formally, this
means that in place of the pre-passage Hermitian lower-case matrix
(\ref{primo}) living, by definition,  in the Hermitian dynamical
regime with $(c,t)\in
 {\cal D}_{(conventional)}$, i.e., with $c \leq t^2 $,
the role of the energies was changed and played by the (strictly
real) eigenvalues of the new, non-Hermitian matrix (\ref{rosecumo}).

The passage from model (\ref{seprimo}) to its phenomenologically
acceptable continuation (\ref{rosecumo}) was postulated smooth.
After one leaves the safe textbook half-plane with $c \leq 0$ and
after one moves to the small and positive $c$'s, one does not
observe any {\em qualitative} changes in the time-dependence of the
energy levels at the instants of transition $t=\pm \sqrt{c}$.
Incidentally, a similar smoothness characterizes the phase
transition occurring in the Lipkin-Meshkov-Glick-type model of
Ref.~\cite{Geyer} (cf. Figure Nr. 2 in {\it loc. cit.}). In our
present model, this smoothness (i.e., the phase transition of the
second kind \cite{Denis}) is illustrated by Fig.~\ref{rija3} where
the smooth spectral shape does not offer any indication that in the
interval of times $t \in (-\sqrt{c},\sqrt{c})$ (indicated by the two
thin markers below the curves) the Hermitian matrix (\ref{primo})
gets replaced by its non-Hermitian continuation (\ref{rosecumo}).

\section{Conclusions}

From the historical perspective it was fortunate that during the
birth of quantum mechanics people did not pay too much attention to
the elementary hydrogen-type quantum systems in which the electrons
(= fermions) would be replaced by pions (i.e., bosons). This would
almost certainly slow down the early stages of development. The
experiments would be found to disagree with the theory and the
theory would suffer from the emergence of multiple theoretical
challenges including, first of al, the manifest non-Hermiticity of
the underlying relativistic Klein-Gordon Hamiltonian (cf., e.g., pp.
pp. 357 -- 360 in \cite{Constantinescu} for more details). Some of
these conceptual questions remained, for a long time, unanswered
(cf., e.g., the related remark on p. 349 in \cite{Constantinescu}).

The much-delayed consequent resolution of the problem was only
published, cca 15 years ago, in Ref.~\cite{aliKG}. An acceptable
probabilistic interpretation of the pionic-atom-like quantum
mechanics was based there, in essence, on the application of the
concept of the quasi-Hermiticity. Unfortunately, the solution of the
Klein-Gordon puzzle was still incomplete, based on the very strong
assumption that the system in question is static (cf., e.g., Theorem
2 in Ref.~\cite{ali}). In other words, for the pionic atoms the
Klein-Gordon Hamiltonians were only shown to describe the critical
quantum phenomena (like, e.g., the complexification of the energies)
in an adiabatic approximation.

The approximation-free 3HS formalism able to provide a complete
description of the processes of quantum degeneracies has only been
formulated very recently (see, e.g., Ref.~\cite{NIP}, with further
references listed therein). In our present paper we imagined that it
would be desirable to apply such a formalism to the evolution of the
systems which happen to pass from the Hermitian to quasi-Hermitian
dynamical regime. Via an elementary illustrative example we
explained the key ideas of the approach and we demonstrated that the
detailed description of an elementary toy model is able to throw
light on multiple conceptual questions. Among them, we made it clear
that

\begin{itemize}

\item
the analysis restricted to the mere description of the energy levels
need not provide any hint that the system is going to pass through a
quantum phase transition;

\item
the traditional textbook versions of quantum theory based on the
work with the fixed physical Hilbert space and with the operators of
observables which are self-adjoint in this space admit extensive
generalizations;

\item
one of these generalizations has been developed and discussed here
via an elementary benchmark model in which the dynamical input
knowledge has the form of a given non-Hermitian and time-dependent
energy operator $H(c,t)$ with real spectrum.

\end{itemize}

 \noindent
At the end, the study led to the rather optimistic conclusion that
the 3HS formalism is able to provide a consistent and mathematically
correct representation of physical reality in which the description
of the passage of a quantum system through its
Hermitian--quasi-Hermitian interface exhibits a number of close
analogies with its simpler, adiabatic-approximation predecessors.

\section*{Acknowledgments}
The work was supported by the GA\v{C}R Grant Nr. 16-22945S.

\end{document}